\input harvmac
\input epsf

\overfullrule=0pt
\abovedisplayskip=12pt plus 3pt minus 3pt
\belowdisplayskip=12pt plus 3pt minus 3pt
%
%
\font\zfont = cmss10 

%

\def\tilde{\widetilde}
\def\bar{\overline}
\def\to{\rightarrow}

\def\cN{{\cal N}}

\def\bigone{\hbox{1\kern -.23em {\rm l}}}
\def\ZZ{\hbox{\zfont Z\kern-.4emZ}}

\font\zfont = cmss10 

\def\bigone{\hbox{1\kern -.23em {\rm l}}}
\def\ZZ{\hbox{\zfont Z\kern-.4emZ}}




\def\frac#1#2{{#1 \over #2}}

\def\cite#1{\#1}

\def\section#1{\newsec {#1}}
\def\subsection#1{\subsec {#1}}
\def\subsubsection#1{\bigskip\noindent{\it #1}}


\def\myI#1 {\int \! #1 \,}

\def\eqn#1#2{\xdef #1{(\secsym\the\meqno)}\writedef{#1\leftbracket#1}%
	$$#2\eqno#1\eqlabeL#1$$%
	\xdef #1{Eq.~(\secsym\the\meqno) }\global\advance\meqno by1}
%



\def\paren#1{\left( #1 \right)}



\def\pa{\partial}

\def\cA{{\cal A}}




%
\let\useblackboard=\iftrue
%
%
\newfam\black
\def\Title#1#2{\rightline{#1}
\ifx\answ\bigans\nopagenumbers\pageno0\vskip1in%
\baselineskip 15pt plus 1pt minus 1pt
\else
\def\listrefs{\footatend\vskip 1in\immediate\closeout\rfile\writestoppt
\baselineskip=14pt\centerline{{\bf References}}\bigskip{\frenchspacing%
\parindent=20pt\escapechar=` \input
refs.tmp\vfill\eject}\nonfrenchspacing}
\pageno1\vskip.8in\fi \centerline{\titlefont #2}\vskip .5in}

\ifx\answ\bigans\def\tcbreak#1{}\else\def\tcbreak#1{\cr&{#1}}\fi
\useblackboard
\message{If you do not have msbm (blackboard bold) fonts,}
\message{change the option at the top of the tex file.}
\font\blackboard=msbm10 scaled \magstep1
\font\blackboards=msbm7
\font\blackboardss=msbm5
\textfont\black=\blackboard
\scriptfont\black=\blackboards
\scriptscriptfont\black=\blackboardss

\else

\fi



%
\message{S-Tables Macro v1.0, ACS, TAMU (RANHELP@VENUS.TAMU.EDU)}
%
%
\newhelp\stablestylehelp{You must choose a style between 0 and 3.}%
\newhelp\stablelinehelp{You should not use special hrules when stretching
a table.}%
\newhelp\stablesmultiplehelp{You have tried to place an S-Table
inside another S-Table.  I would recommend not going on.}%
%
%
\newdimen\stablesthinline
\stablesthinline=0.4pt
\newdimen\stablesthickline
\stablesthickline=1pt
%
%
\newif\ifstablesborderthin
\stablesborderthinfalse
\newif\ifstablesinternalthin
\stablesinternalthintrue
\newif\ifstablesomit
\newif\ifstablemode
\newif\ifstablesright
\stablesrightfalse
%
%
\newdimen\stablesbaselineskip
\newdimen\stableslineskip
\newdimen\stableslineskiplimit
%
%
\newcount\stablesmode
\newcount\stableslines
\newcount\stablestemp
\stablestemp=3
\newcount\stablescount
\stablescount=0
\newcount\stableslinet
\stableslinet=0
%
%
%
\newcount\stablestyle
\stablestyle=0
%
%
\def\stablesleft{\quad\hfil}%
\def\stablesright{\hfil\quad}%
%
%
\catcode`\|=\active%
%
%
\newcount\stablestrutsize
\newbox\stablestrutbox
\setbox\stablestrutbox=\hbox{\vrule height10pt depth5pt width0pt}
\def\stablestrut{\relax\ifmmode%
                         \copy\stablestrutbox%
                       \else%
                         \unhcopy\stablestrutbox%
                       \fi}%
%
%
\newdimen\stablesborderwidth
\newdimen\stablesinternalwidth
\newdimen\stablesdummy
\newcount\stablesdummyc
\newif\ifstablesin
\stablesinfalse
%
%
%
%
%
\def\stablesadj{%
  \ifcase\stablestyle%
    \hbox to \hsize\bgroup\hss\vbox\bgroup%
  \or%
    \hbox to \hsize\bgroup\vbox\bgroup%
  \or%
    \hbox to \hsize\bgroup\hss\vbox\bgroup%
  \or%
    \hbox\bgroup\vbox\bgroup%
  \else%
    \errhelp=\stablestylehelp%
    \errmessage{Invalid style selected, using default}%
    \hbox to \hsize\bgroup\hss\vbox\bgroup%
  \fi}%
\def\stablesend{\egroup%
  \ifcase\stablestyle%
    \hss\egroup%
  \or%
    \hss\egroup%
  \or%
    \egroup%
  \or%
    \egroup%
  \else%
    \hss\egroup%
  \fi}%
\def\stablestart{%
  \ifstablesin%
    \errhelp=\stablesmultiplehelp%
    \errmessage{An S-Table cannot be placed within an S-Table!}%
  \fi
  \global\stablesintrue%
  \global\advance\stablescount by 1%
  \message{<S-Tables Generating Table \number\stablescount}%
  \begingroup%
  \stablestrutsize=\ht\stablestrutbox%
  \advance\stablestrutsize by \dp\stablestrutbox%
  \ifstablesborderthin%
    \stablesborderwidth=\stablesthinline%
  \else%
    \stablesborderwidth=\stablesthickline%
  \fi%
  \ifstablesinternalthin%
    \stablesinternalwidth=\stablesthinline%
  \else%
    \stablesinternalwidth=\stablesthickline%
  \fi%
  \tabskip=0pt%
  \stablesbaselineskip=\baselineskip%
  \stableslineskip=\lineskip%
  \stableslineskiplimit=\lineskiplimit%
  \offinterlineskip%
  \def\borderrule{\vrule width \stablesborderwidth}%
  \def\internalrule{\vrule width \stablesinternalwidth}%
  \def\thinline{\noalign{\hrule height \stablesthinline}}%
  \def\thickline{\noalign{\hrule height \stablesthickline}}%
  \def\trule{\omit\leaders\hrule height \stablesthinline\hfill}%
  \def\ttrule{\omit\leaders\hrule height \stablesthickline\hfill}%
  \def\tttrule##1{\omit\leaders\hrule height ##1\hfill}%
  \def\stablesel{&\omit\global\stablesmode=0%
    \global\advance\stableslines by 1\borderrule\hfil\cr}%
  \def\el{\stablesel&}%
  \def\elt{\stablesel\thinline&}%
  \def\eltt{\stablesel\thickline&}%
  \def\elttt##1{\stablesel\noalign{\hrule height ##1}&}%
  \def\elspec{&\omit\hfil\borderrule\cr\omit\borderrule&%
              \ifstablemode%
              \else%
                \errhelp=\stablelinehelp%
                \errmessage{Special ruling will not display properly}%
              \fi}%
  \def\stmultispan##1{\mscount=##1 \loop\ifnum\mscount>3 \stspan\repeat}%
  \def\stspan{\span\omit \advance\mscount by -1}%
  \def\multicolumn##1{\omit\multiply\stablestemp by ##1%
     \stmultispan{\stablestemp}%
     \advance\stablesmode by ##1%
     \advance\stablesmode by -1%
     \stablestemp=3}%
  \def\multirow##1{\stablesdummyc=##1\parindent=0pt\setbox0\hbox\bgroup%
    \aftergroup\emultirow\let\temp=}
  \def\emultirow{\setbox1\vbox to\stablesdummyc\stablestrutsize%
    {\hsize\wd0\vfil\box0\vfil}%
    \ht1=\ht\stablestrutbox%
    \dp1=\dp\stablestrutbox%
    \box1}%
  \def\stpar##1{\vtop\bgroup\hsize ##1%
     \baselineskip=\stablesbaselineskip%
     \lineskip=\stableslineskip%
   \lineskiplimit=\stableslineskiplimit\bgroup\aftergroup\estpar\let\temp=}%
  \def\estpar{\vskip 6pt\egroup}%
  \def\stparrow##1##2{\stablesdummy=##2%
     \setbox0=\vtop to ##1\stablestrutsize\bgroup%
     \hsize\stablesdummy%
     \baselineskip=\stablesbaselineskip%
     \lineskip=\stableslineskip%
     \lineskiplimit=\stableslineskiplimit%
     \bgroup\vfil\aftergroup\estparrow%
     \let\temp=}%
  \def\estparrow{\vfil\egroup%
     \ht0=\ht\stablestrutbox%
     \dp0=\dp\stablestrutbox%
     \wd0=\stablesdummy%
     \box0}%
  \def|{\global\advance\stablesmode by 1&&&}%
  \def\|{\global\advance\stablesmode by 1&\omit\vrule width 0pt%
         \hfil&&}%
\def\vt{\global\advance\stablesmode
by 1&\omit\vrule width \stablesthinline%
          \hfil&&}%
  \def\vtt{\global\advance\stablesmode by 1&\omit\vrule width
\stablesthickline%
          \hfil&&}%
  \def\vttt##1{\global\advance\stablesmode by 1&\omit\vrule width ##1%
          \hfil&&}%
  \def\vtr{\global\advance\stablesmode by 1&\omit\hfil\vrule width%
           \stablesthinline&&}%
  \def\vttr{\global\advance\stablesmode by 1&\omit\hfil\vrule width%
            \stablesthickline&&}%
\def\vtttr##1{\global\advance\stablesmode
 by 1&\omit\hfil\vrule width ##1&&}%
  \stableslines=0%
  \stablesomitfalse}
\def\stablesdef{\bgroup\stablestrut\borderrule##\tabskip=0pt plus 1fil%
  &\stablesleft##\stablesright%
  &##\ifstablesright\hfill\fi\internalrule\ifstablesright\else\hfill\fi%
  \tabskip 0pt&&##\hfil\tabskip=0pt plus 1fil%
  &\stablesleft##\stablesright%
  &##\ifstablesright\hfill\fi\internalrule\ifstablesright\else\hfill\fi%
  \tabskip=0pt\cr%
  \ifstablesborderthin%
    \thinline%
  \else%
    \thickline%
  \fi&%
}%
\def\endtable{\advance\stableslines by 1\advance\stablesmode by 1%
   \message{- Rows: \number\stableslines, Columns:  \number\stablesmode>}%
   \stablesel%
   \ifstablesborderthin%
     \thinline%
   \else%
     \thickline%
   \fi%
   \egroup\stablesend%
\endgroup%
\global\stablesinfalse}
%


\def\em {\it}


\Title{\vtop{\hbox{hep-th/0112064}
\hbox{SU-ITP-01/51}}}
{\vbox{\centerline{Noncommutative Dipole Field Theories}}}
\centerline{ K. Dasgupta\foot{\tt keshav@itp.stanford.edu},
~M. M. Sheikh-Jabbari\foot{\tt jabbari@itp.stanford.edu}}
\vskip 15pt
\centerline{\it Department of Physics, Stanford University}
\centerline{\it 382 via Pueblo Mall, Stanford CA 94305-4060, USA}

\ \smallskip
\centerline{\bf Abstract}

Assigning an intrinsic constant dipole moment to any field, we present a
new kind of associative star product, the dipole star product, which
was first introduced in [hep-th/0008030]. We develop the mathematics
necessary to study the corresponding noncommutative dipole field theories.
These theories  are sensible non-local
field theories with no IR/UV mixing.
In addition we discuss that the Lorentz symmetry in these theories is
``softly''
broken and in some particular  cases the CP (and even CPT) violation in these
theories  may become observable.
We show that a non-trivial dipole extension of $\cN=4$, $D=4$ gauge
theories can only be obtained if we break the $SU(4)$ R (and hence
super)-symmetry. Such noncommutative dipole extensions, which in the
maximal supersymmetric cases are $\cN=2$ gauge theories with matter, can
be embedded in string theory as the theories on  D3-branes probing a
smooth Taub-NUT space with three form fluxes
turned on or alternatively by probing a space with R-symmetry twists. We show
the equivalences between the two approaches and also discuss the M-theory
realization.

\Date{December 2001}
{\vfill\eject}
\ftno=0

\lref\Witt{E. Witten,
  {\it ``Bound States Of Strings And $p$-Branes,''}
Nucl.Phys. {\bf B460} (1996) 335, hep-th/9510135.}

\lref\CDS{A.~Connes, M.R.~Douglas and A.~Schwarz,
  {\it ``Noncommutative Geometry and Matrix Theory:
  Compactification on Tori,''} JHEP {\bf 9802} (1998) 003,
hep-th/9711162.}

\lref\DH{M.R.~Douglas and C.~Hull,
  {\it ``D-branes and the Noncommutative Torus,''}
  JHEP {\bf 9802} (1998) 008, hep-th/9711165.}

\lref\SWNCG{N. Seiberg and E. Witten,
  {\it ``String Theory and Noncommutative Geometry,''}
  JHEP {\bf 9909} (1999) 032, hep-th/9908142.}

\lref\DIP{M.M. Sheikh-Jabbari,
  {\it ``
 Open Strings in a B-field Background as Electric Dipoles,''}
 Phys. Lett. {\bf B455} (1999) 129, hep-th/9901080.}

\lref\Suss{D. Bigatti, L. Susskind,
  {\it ``
 Magnetic Fields, Branes and Noncommutative Geometry,''}
Phys.Rev. {\bf D62} (2000) 066004, hep-th/9908056.}

\lref\Motl{ L. Motl,
  {\it ``
Melvin Matrix Models,''} hep-th/0107002.}

\lref\Ren{M.M. Sheikh-Jabbari,
  {\it ``One Loop Renormalizability of Supersymmetric Yang-Mills Theories on
Noncommutative Two-Torus,''} JHEP {\bf 9906} (1999) 015,
hep-th/9903107.}

\lref\MRS{S. Minwalla, M. Van Raamsdonk, N. Seiberg
  {\it ``Noncommutative Perturbative Dynamics,''}
JHEP {\bf 0002} (2000) 020, hep-th/9912072.}

\lref\GMS{R. Gopakumar, S. Minwalla, A. Strominger,
  {\it ``Noncommutative Solitons,''}
JHEP {\bf 0005} (2000) 020, hep-th/0003160.}

\lref\AOI{M. Alishahiha, H. Ita, Y. Oz,
{\it ``Graviton Scattering on D6 Branes with B Fields,''}
JHEP {\bf 0006} (2000) 002, hep-th/0004011.}

\lref\Haya{M.Hayakawa,
  {\it ``Perturbative Analysis on Infrared and Ultraviolet Aspects of
Noncommutative QED on $R^4$,''} hep-th/ 99121167.}

\lref\NCSM{M.~Chaichian, P.~Presnajder, M.~M.~Sheikh-Jabbari and A.~Tureanu,
{\it ``Noncommutative standard model: Model building,''}
hep-th/0107055.}

\lref\CPT{M.M. Sheikh-Jabbari,
  {\it ``Discrete Symmetries (C,P,T) in Noncommutative Field Theories,''}
Phys.Rev. Lett. {\bf 84} (2000) 5265, hep-th/ 0001167.}

\lref\Harvey{ J. A. Harvey,
  {\it ``Magnetic Monopoles, Duality, and Supersymmetry,''}
hep-th/ 9603086.}

\lref\aki{A.~Hashimoto and N.~Itzhaki,
{\it ``Noncommutative Yang-Mills and the AdS/CFT Correspondence,''}
Phys.\ Lett.\ B {\bf 465}, 142 (1999),
hep-th/9907166.}

\lref\juan{J.~M.~Maldacena and J.~G.~Russo,
{\it ``Large N Limit of Noncommutative Gauge Theories,''}
JHEP {\bf 9909}, 025 (1999),
hep-th/9908134.}

\lref\BerGan{A. Bergman and O.J. Ganor,
  {\it ``Dipoles, Twists and Noncommutative Gauge Theory,''}
JHEP {\bf 0010}, 018 (2000),
  hep-th/ 0008030.}

\lref\CDGR{S. Chakravarty, K. Dasgupta,
   O.J. Ganor and G. Rajesh,
    {\it ``Pinned Branes and New Non-Lorentz Invariant Theories,''}
   Nucl.Phys. {\bf B587} (2000) 228, hep-th/ 0002175.}

\lref\bkar{D.~S.~Bak and A.~Karch,
{\it ``Supersymmetric brane-antibrane configurations,''}
hep-th/0110039.}

\lref\Ken{K. Intriligator,
  {\it ``Maximally Supersymmetric RG Flows and AdS Duality,''}
Nucl.Phys. {\bf B580} (2000) 99,
hep-th/9909082.}

\lref\Jackson{ J. D. Jackson {\it `` Classical Electro-Dynamics,''}
New York : Wiley, c1999.}

\lref\ParticleData{ D. E. Groom, {\it et. al., ``The Review of
Particle Physics,''} Eur. Phys. J {\bf C15} (2000) 1.}

\lref\HASI {A.~Hashimoto and N.~Itzhaki,
  {\it ``Noncommutative Yang-Mills and the AdS/CFT correspondence,''}
  Phys. Lett. {\bf B465} (1999) 142,
  hep-th/9907166.}

\lref\MALRUS{
  J.~M.~Maldacena and J.~G.~Russo,
  {\it ``Large N limit of Noncommutative Gauge Theories,''}
  JHEP {\bf 9909} (1999) 025, hep-th/9908134.}

\lref\ZYIN{Z. Yin, {\it ``A Note on Space Noncommutativity,''}
   Phys. Lett. {\bf B466} (1999) 234, hep-th/9908152.}

\lref\DGR {K. Dasgupta, O. J. Ganor and G. Rajesh,
  {\it ``Vector Deformations of $\cN = 4$ Yang-Mills Theory, Pinned Branes
 and Arched Strings,''} hep-th/0010072.}

\lref\DolNap {L. Dolan and C. Nappi,
  {\it ``A Scaling Limit with Many Noncommutativity Parameters,''}
   hep-th/0009225.}

\lref\RT {R. Tatar, {\it ``A Note on Noncommutative Field Theory and
  Stability of Brane- Antibrane Systems,''}
 hep-th/0009213.}

\lref\WITSFT {E. Witten, {\it ``Noncommutative Geometry and String Field
  Theory,''} Nucl. Phys. {\bf B268} (1986) 253.}

\lref\DGRunp {K. Dasgupta, O. J. Ganor and G. Rajesh, {\it unpublished}.}

\lref\CHU{C.-S. Chu and P.-M. Ho, {\it ``Constrained Quantization of Open
 String in Background $B$-Field and Noncommutative D-Brane,''}
Nucl. Phys. {\bf B568} (2000) 447, hep-th/9906192.}

\lref\KONTS {M. Kontsevich, {\it ``Deformation Quantization of Poisson
 Manifolds,''} q-alg/9709040.}

\lref\CandF {A. S. Cattaneo and  G. Felder, {\it ``A Path Integral Approach
 to Kontsevich Quantization Formula,''}
  Commun. Math. Phys. {\bf 212} (2000) 591, math.qa/9902090.}

\lref\DGS{ K. Dasgupta, G. Rajesh and S. Sethi, {\it ``M-Theory,
 Orientifolds and G-Flux,''} JHEP {\bf 9908} (1999) 023, hep-th/9908088.}

\lref\SJ{F. Ardalan, H. Arfaei and  M.M. Sheikh-Jabbari, {\it ``Dirac
Quantization of Open
 Strings and Noncommutativity in Branes,''}
 Nucl. Phys. {\bf B576} (2000) 578, hep-th/9906161;
 M.M. Sheikh-Jabbari and A. Shirzad, {\it ``Boundary Conditions as Dirac
 Constraints,''}
Eur.Phys.J. {\bf C19} (2001) 383, hep-th/9907055.}

\lref\future{K. Dasgupta and M.M. Sheikh-Jabbari, {\it ``Work in
Progress''}.}

\lref\KesYin{K. Dasgupta and Z. Yin, {\it ``Non-{\cA}belian Geometry,''}
hep-th/0011034.}

\lref\Arefeva{ I. Ya. Aref'eva, D. M. Belov, A. S.Koshelev,
{\it ``A Note on UV/IR for Noncommutative Complex Scalar Field,''}
hep-th/0001215.}

\lref\DMUK{K. Dasgupta and S. Mukhi, {\it ``Brane Constructions, Fractional
Branes and Anti- deSitter Domain Walls,''} JHEP {\bf 9907} (1999) 008,
hep-th/9904131.}

\lref\BDGKR{A. Bergman, K. Dasgupta, O. J. Ganor, J. L. Karczmarek,
G. Rajesh, {\it ``Non-local Field Theories and Their Gravity Duals,''}
hep-th/0103090.}

\section{Introduction}
In the last three years a great amount of work have been devoted to the
field theories on the Moyal plane, usually called noncommutative field
theories. The noncommutative Moyal plane defined by the coordinate
operators with
\eqn\Moyal{ [x^\mu, x^\nu]=i\theta^{\mu\nu}\ ,}
where  $\theta$ is a constant, can be realized in string theory as the
world volume of D-branes in a constant background $B_{\mu\nu}$ field,
probed by the open strings \refs{\SJ , \CHU , \SWNCG}. In the absence of
D-branes, a constant B-field can be gauged away completely.
However, in the
presence of D-branes constant $B_{\mu\nu}$ fields with both $\mu$ and
$\nu$ directions  along the brane cannot be gauged away \refs{\Witt} and in
fact such components lead to the noncommutativity on the brane world
volume.
The effective low energy world volume field theories on the brane
turn out to be noncommutative (supersymmetric) gauge theories.
The very characteristic of noncommutative gauge theories, in general, is
that they are non-local theories with the fields effectively describing
the dynamics of ``dipole'' like objects \refs{\DIP, \Ren}.
The dipole moments of these particles are proportional to their momentum
\eqn\ncdip{
d_{\mu}\sim \theta_{\mu\nu} p_{\nu}\ .}
This dipole nature is the familiar effect of the motion of particles in
an external magnetic field \refs{\Suss }.
{}From field theory point of view the momentum dependence of the  dipole
moments (and the corresponding non-locality) shows up in the loop
expansion of the noncommutative field theories as the IR/UV mixing
\refs{\MRS}.

The components of the B-field not parallel to the D-branes however can be
gauged away if the B-field is a constant. This
can be waived if in some way we stabilize the B-field along
the transverse directions to the D-brane to support a non-constant
B-flux.
This is possible, for example, if we compactify the transverse
direction to a brane with a varying size of the compact circle. Putting one
leg of the B-field along that direction effectively gives us a non-zero
three form $H_{NS}$ field.
The case that we want to study further in
this paper, however,
 is to stabilize the B-field when it has one leg along the brane
and the other transverse to it. The possibility of such configuration was
discussed earlier in \refs{\CDGR , \BerGan , \DGR }, where
the ``twisted'' compactification were introduced.
We shall review this in section five.
As discussed in \refs{\BDGKR} performing the ``twisted'' compactification
leads us to introduce a new type of star product between the fields
at the level of effective field theories. As a result of the twisted
compactification we find the possibility of associating a {\it
constant}  dipole  length to any field. This dipole
length is proportional to the ``winding'' number of
the fields in the twisted directions. We shall review this in section five.
Then, at the level of the effective
field theory we obtain some sort of noncommutative field theory which
we call  ``noncommutative dipole field theory'' (NCDFT) to distinguish
it from the theories on the Moyal plane. We would like to stress that unlike
the Moyal case, the origin of the noncommutativity in noncommutative
dipole field theories is not the noncommutativity in space-time, but an
inherent property of each field and we can also have fields with zero dipole
length.

Besides the string theoretic appeal, the
noncommutative dipole field theories, NCDFT's,  are also interesting by
themselves. As we will discuss, considering some definite noncommutative
dipole gauge theories, there is the chance of finding a
CP (and even CPT) violating theory. We also discuss the renormalizability of
noncommutative dipole
scalar and gauge field theories and argue that these theories are
{\it renormalizable} in the sense that adding finite numbers of
{\it non-local} counterterms at each loop level will remove the divergences.
We show that the $\beta$-function of the noncommutative
dipole QED is not affected by the dipole nature
of the fields. Furthermore we argue that, in general,  there is no IR/UV
mixing
effect in the noncommutative dipole theories.
We show that in the noncommutative gauge theory with a fundamental matter
fields, unlike the Moyal case, we can have $SU(N)$ gauge theories.
However, for the adjoint matter field, we will argue that $SU(N)$ is not
possible and one should take  $U(N)$.

We then study the possible supersymmetric extension of the
noncommutative dipole theories.
We show that unlike the Moyal noncommutative gauge theories, the maximal SUSY
noncommutative dipole gauge theory is the $\cN=2$ (eight supercharges).

The plan of this paper is as follows. In section 2, first we present the
explicit form of the  dipole star product and study some of its
properties, such as associativity. We show that the usual integral
over the space-time can provide a natural $Tr$ over the $C^*$-algebra of
functions with the dipole star products.  Then in section 3, we study
noncommutative dipole field theories and their renormalizability.
In section 4, we discuss the SUSY extension of noncommutative dipole gauge
theories. In section 5, the string theory realization of NCDFT's
is discussed (the first two parts of this section is  an expanded review
of Ref.\refs{\BDGKR}). We also discuss how the NCDFT's can be understood as
the effective field theories on D3-branes probing a Taub-NUT space with a
B-field switched on.
Section 6 is devoted to the M theory realizations.
In section 7 we study another kind of star-product which uses both the
dipole nature of open-strings and the noncommutativity of the underlying
space-time.
We end with a discussion and outlook.

\section{Mathematical Preliminaries}

To formulate any noncommutative field theory, one should start by defining
 the proper {\it associative} star
product. Then, the  fields are members of the $C^*$-algebra of functions
with respect to that star product. To any element of the  $C^*$-algebra,
$\phi_i$, we assign a {\it constant} space-like dipole length $\vec{L_i}$
and define the ``dipole star product'' as
\eqn\prodrule{
(\phi_i * \phi_j)(x) \equiv \phi_i(x - {1\over 2} L_j)~ \phi_j (x +
{1\over 2}L_i)\ .}
Furthermore we need to identify the dipole moment of $(\phi_i *
\phi_j)(x)$. Demanding our algebra to be associative, it is
straightforward to show that {\it the dipole moment of (star) product of
two functions should be sum of their dipole moments}:
\eqn\associate{\eqalign{
(\phi_i * \phi_j)*\phi_k &= \big(\phi_i(x - {1\over 2} L_j)\ \phi_j (x
+{1\over 2}L_i)\big) *\phi_k \cr
&=  \phi_i(x - {L_j+L_k \over 2})\ \phi_j (x +{L_i-L_k\over 2})\
\phi_k(x+{L_i+L_j\over 2})\cr &= \phi_i(x-{L_j+L_k\over 2})\ (\phi_j*
\phi_k)(x +{1\over 2}L_i)\cr &= \phi_i * (\phi_j *\phi_k)}\ .}
We would like to mention that the above star product cannot be expressed
in terms of a commutation relation among the space-time coordinates.
In other words, the noncommutativity in the dipole
case is not a property of space-time, but originating from the dipole
length associated to each field.

Now we should introduce a suitable $Tr$ on the algebra. The
natural choice for the $Tr$, similar to the Moyal case, is the integral
over the space-time. However naively taking the integral over star
products
of arbitrary functions, one can easily check that they do not enjoy
the necessary cyclicity condition. This problem is removed if we
restrict the integrand to have a total zero dipole length. More
explicitly, the
integral serves as the proper $Tr$, over the functions of zero dipole
length:

\eqn\intergal{
\int \phi_1 *\phi_2 *\cdots * \phi_n = \int \phi_n* \phi_1 *\cdots
\phi_{n-1}\ , }
with the condition that
$\sum_{i=1}^n \vec{L_i}=0$, where $L_i$ are the dipole lengths for
$\phi_i$.
In the field theory actions this condition is translated into
the
fact that any term in the proposed action should have a total vanishing
dipole length (as well as the usual hermiticity conditions which
guarantees the electric charge conservation). Therefore, maintaining with
the translational invariance, in each vertex both
the sum of the external momenta and the total dipole length
should vanish.

Next, we should define complex conjugate of a field and the behaviour of
the star product under complex conjugation. Demanding
$(\phi^{\dagger}*\phi)$ to be real valued, i.e.
\eqn\phidagger{
 \phi^\dagger *\phi=(\phi^\dagger\ *\phi)^{\dagger}\ , }
fixes the dipole length of $\phi^\dagger$ to be the same as that of $\phi$
though with a minus sign. Therefore the dipole length of any real
(hermitian) field, and in particular the gauge fields, is zero.
Then, one can show that
\eqn\complex{
(\phi_i * \phi_j)^{\dagger}= \phi_j^{\dagger}*\phi_i^{\dagger}}

To write down the field theory action we also need the derivative
operators with respect to the star product. It is easy to check that the
usual derivative does satisfy the Leibniz rule, i.e.
\eqn\Leib{
\partial_\mu (\phi_1 *\phi_2)= (\partial_\mu \phi_1) *\phi_2+
\phi_1 * \partial_\mu \phi_2\ ,}
i.e. they can be used as the proper derivatives for writing down the
kinetic terms of the NCDFT's.

\section{Noncommutative Dipole Field Theories}

Equipped with the above mathematics we are ready to formulate
noncommutative dipole field theories (NCDFT's).
In general, to obtain the NCDFT's actions it is
enough to replace the product of fields in the commutative actions with
the dipole star product \prodrule. However, one should insert the proper
dipole lengths for all fields.
We start with a simple scalar field theory and then formulate
fermions and gauge theories.
\subsection{Scalar noncommutative dipole theory}

Along our general recipe, the action proposed for the scalar
noncommutative dipole theory is
\eqn\actionscalar{
S=\int \partial_\mu\phi^\dagger * \partial_\mu\phi - V_*(\phi^\dagger
*\phi)\ ,}
where $V_*$ is the potential with the products replaced by star products
\prodrule. We note that, written as a function of $\phi^\dagger *
\phi$, it is guaranteed that  the terms in the above action satisfy the
necessary zero-dipole-length condition. Using the cyclicity condition,
we can drop the star product in the quadratic part of the action(s) (much
like the Moyal-star product case) and the effects of dipole moments
appear only through the interaction terms.  Therefore the propagator
of the above noncommutative dipole theory (and also any NCDFT) is the same
as the commutative case.
Moreover, since
$$
(\phi^\dagger *\phi)(x)=(\phi^\dagger \phi)(x-{1\over 2}L)\ ,
$$
then
\eqn\potential{\eqalign{
\int (\phi^\dagger *\phi)_*^n &=\int \left((\phi^\dagger \phi)(x-{1\over
2}L)\right)^n \cr
&=\int \left((\phi^\dagger \phi)(x)\right)^n\ . }}
Hence, even in the potential term, being only a function of $\phi^\dagger
*\phi$, all the dipole dependence is removed.
As a result the scalar noncommutative dipole theory introduced by the
action \actionscalar is exactly the same as its commutative counter-part.

Noting that $\phi *\phi^{\dagger}$ is also of zero dipole length,
there is another possibility for potential: to be a function of
both $\phi *\phi^{\dagger}$ and $\phi^{\dagger} *\phi$. Hence,
potential for the noncommutative dipole version of the
$\phi^4$ theory in the most general case is
\eqn\potential{
V_*=\lambda_0(\phi^\dagger *\phi) *(\phi^\dagger *\phi)
+\lambda_1(\phi^\dagger *\phi) *(\phi *\phi^\dagger)\ ,}
and therefore
\eqn\potint{
\int V_*=
\int \lambda_0 (\phi^\dagger \phi)(x)(\phi^\dagger \phi)(x)+
\lambda_1 (\phi^\dagger \phi)(x - {L\over 2})(\phi^\dagger \phi)(x +
{L\over 2}).}

Performing loop calculations, one can show that at the level of one loop,
two point functions will show the same kind of divergences as the
commutative
theory. Furthermore, we have both planar
and non-planar diagrams (where the exponential phases involving the dipole
length and the loop momenta appear in the loop integrals). The
non-planar diagrams are finite and  we do not face IR/UV mixing, because
the dipole length, unlike the Moyal noncommutative case, is a constant.
However, considering the one loop four point function, something interesting
happens now. We have planar and non-planar one loop four point functions
and again non-planar diagrams are finite. As for the planar diagrams,
we face two type of divergences: those
which can be cancelled with the usual counter-terms {\it if}
$\lambda_0=\lambda_1$, and those which cannot be absorbed in the terms
already present in the action. Therefore, the noncommutative dipole theory
with potential \potential is {\it not} renormalizable in the usual sense
used for local field theories.
However, these extra divergences can be absorbed in a term like
$$ \int (\phi^\dagger \phi)(x - {2L\over 2})(\phi^\dagger \phi)(x +
{2L\over 2})\ .
$$
This procedure should be continued to the $m$ loops order.
In the loops of order $m$ the divergences in this theory can be
cancelled by the following choice of counter-terms:
\eqn\counterterm{
\sum_{n=0}^{m+1}~\lambda_n \int (\phi^\dagger \phi)(x - {nL\over
2})(\phi^\dagger \phi)
(x + {n L\over 2}).}

Although the above noncommutative dipole theory is not
renormalizable in the usual sense, we see that the
counter-terms are under strict control $-$ in fact are of the same form as
the original Lagrangian $-$ and we do not require any extra degrees of
freedom at the UV.
\foot{
We would like to comment that  for the Moyal noncommutative case
again we have two possibilities for strict renormalizabilty in the usual sense.
Besides
$\lambda_1=0$, the $\lambda_0=\lambda_1$ case is also renormalizable
\refs{\Arefeva} with $\lambda_i=0$ for $i \ge 2$.}
In particular if instead of the potential \potential we start
with \counterterm (with the sum going to infinity) with all the
coefficient $\lambda_n$ to be equal, the theory, though non-local, would
be renormalizable in the usual sense. In fact, we will show in section 4
that, this special case is what one finds from string theory.


\subsection{Noncommutative dipole gauge theory}

Here we restrict ourselves to the $SU(N)$ gauge theories and the other
gauge groups can be studied in the same spirit. As we discussed the gauge
fields, being hermitian, should have a zero dipole length. So, the pure
gauge theory is defined exactly in the same way as the commutative gauge
theory.
However, the gauge fields coupled to the matter fields may uncover the
dipole structure of the charged matter fields.  In order to define the
fundamental matter coupled to gauge
field theory, we need the covariant derivative of a dipole field. This is
given as:
\eqn\covariant{\eqalign{
D_\mu\psi &\equiv \partial_\mu \psi+ig A_\mu *\psi\ \cr
&= \partial_\mu \psi+ig A_\mu(x-{L\over 2}) \psi(x)\ .}}
Then it is straightforward to check that the action
\eqn\dipoledirac{
S=\int \bar{\psi} \gamma^\mu D_\mu\psi \ ,}
is invariant under gauge transformations:
\eqn\gaugetrans{\eqalign{
\psi &\to U*\psi\cr
A_\mu &\to U*A_\mu*U^{-1}+{i\over g}\partial_\mu U
*U^{-1}=UA_\mu\ U^{-1}+{i\over g}\partial_\mu U U^{-1}\ ,}}
where  $U\in SU(N)$ (and of course the dipole length assigned to $U$
is zero). \foot{Similar to the Moyal noncommutative case \refs{\Haya,
\CPT} we can have
another type of fermion (or covariant derivative):
$$
D_\mu\psi'= \partial_\mu \psi'+ig \psi' * A_{\mu}
= \partial_\mu \psi'+ig A_\mu(x+{L\over 2}) \psi'(x)\ .
$$
The $\psi$ and $\psi'$ type fermions are related by parity transformation,
while  the two types of  fermions in the Moyal NCQED are related by charge
conjugation \refs{\CPT}. The gauge transformation for the $\psi'$ type
fermion is
$\psi'\to \psi' * U^{-1}$.} Note that all the fermions in the $N$-vector
of $SU(N)$ fundamental representation should have the same dipole length.
We would like to comment that, unlike the noncommutative Moyal QED
\refs\Haya, in the
noncommutative dipole QED we can have particles with arbitrary electric
charge.
Expanding \covariant in powers of $L$ the first order terms will give the
dipole interactions, where the dipole moment is
\eqn\dipole{
\vec{d}={1\over 2}g\langle\bar\psi\gamma^0\psi\rangle \vec{L}\ .}

It is worth noting that the noncommutative dipole gauge theory under the
parity is not invariant and as expected the matter field with dipole
length $\vec{L}$ is mapped into the theory with dipole length $-\vec{L}$,
while
under charge conjugation and also time reversal the theory remains
invariant. \foot{
The parity, charge conjugation and time reversal are defined in the same
way as the usual commutative theories.}
So, the dipole theory, with the dipole lengths $\vec{L}_i$ under
CP (as well as CPT) is mapped to another dipole theory with $-\vec{L}_i$.
\foot{ Also note that the in the noncommutative dipole theories
Lorentz symmetry is ``softly''
broken.}
However, we will argue momentarily that this CP and CPT violation is only
observable in very particular cases.
We would also  like to comment that the dipole moment \dipole under both
parity and charge conjugation change sign and hence invariant under CP.

Along our previous discussions the propagator for fermions and the
gauge fields in the dipole case is the same as the commutative case.
However, as it is seen from \covariant the interaction vertices of  the
noncommutative dipole
gauge theory (with ``fundamental'' matter) compared to the commutative
case, involve an extra  dipole
dependent phase factor, $e^{-{i\over 2}p\cdot L}$
(where $p$ is the in-going gauge field momentum).
Performing loop
calculations, since only the momentum of the gauge field appears in the
noncommutative phase factors,  the noncommutative dipole phase
factors never contain the momentum running in the loop. Therefore, we do
not face any non-planar diagram in our noncommutative dipole theory
coupled to the fundamental matter and hence, there is no
IR/UV mixing phenomenon. In particular, the fermion and gauge particle one
loop two point functions are not altered by the dipole length at all.
As for the fermion-gluon vertex,
the noncommutative dipole phase factor will appear just in front of the
loop integrals. Consequently, all the loop integrals are the same as the
commutative case and hence the $\beta$-function of the field theory is the
same as the commutative case. Furthermore, the dipole vector $\vec{L}$ do
not receive any quantum corrections.

It is straightforward to check that
the noncommutative dipole gauge theory is invariant under BRST symmetry.
The corresponding BRST transformations are obtained by inserting the
dipole star product \prodrule into  the commutative expressions. The
ghost field and the BRST generators have zero dipole lengths. Therefore,
the above one loop argument ensures the renormalizability of the theory.

One may wonder now that
the dipole length $L$ is not appearing in the loop
dynamics of the gauge theory, it may be removed by a field re-definition
as:
\eqn\redefine{
A_\mu(x-{L\over 2})=\tilde{A_\mu}(x)\ .}
Then it is easy to check that the theory written in terms of $\tilde A$ is
exactly the usual gauge theory. This field re-definition can be understood
in a more intuitive way. As we know \refs\Jackson , by a translation  in
the origin of the coordinate system, any point like  $2^n$-pole looks as
a bunch of $2^m$-poles $(m\geq n)$. In particular our charged-dipole like
particle can be viewed as a pure charge in a translated frame. This is
just equivalent to the above field re-definition.

Note that the above field re-definition can remove the dipole length $L$
when we have only one kind of particle. If we allow particles of
different dipole lengths, this simple field re-definition will not work.
However, it is not hard to see that, if we have a gauge theory of a {\it
simple} group the CP (and CPT) violating phase factor, even if we have
fundamental matter fields with different dipole lengths, is {\it not}
observable at tree level.
As an explicit example consider the ``noncommutative dipole'' QED with
electron and muon which have dipole lengths $L_e$ and $L_\mu$
respectively. It is straightforward to check that our renormalizability
argument is not affected. Now consider the $e-\mu$ scattering, the
scattering amplitude picks up a phase factor of $e^{ip\cdot(L_e-L_\mu)}$
where $p$ is the momentum of the exchanged photon. Although we have a
non-trivial dipole phase  factor, it will not appear in the cross
sections, i.e. having different dipole lengths is not enough to make the
dipole effects traceable. \foot{We should mention that this is not the case
if we consider loop effects. In particular consider the noncommutative
dipole version of the Hydrogen atom with $L_e$ and $L_P$ for electron and
proton dipoles, respectively.  The $L_e-L_P$ will appear in the potential
term (which can be thought of summing the whole ladder) in the corresponding
Schr\"odinger equation and this will change the
spectrum.} In order to observe noncommutative dipole
phase (for
fundamental matter) in the tree level cross  sections, we need to fulfill two
other
conditions:
(1) We should take a semi-simple group, and
(2) we should allow
particles to have different dipole lengths for different simple group
factors.

These conditions can easily be satisfied in a noncommutative dipole
version of the usual Standard Model, and then the CP and CPT violating
phase factor could in principle be observable in the $e-\mu$
scattering.

\subsection{Adjoint noncommutative dipole matter fields}

Besides the matter fields in the fundamental representation one can
introduce the matter field in the
adjoint representation, with the covariant derivative defined as
\eqn\adjoint{\eqalign{
D_{\mu}\phi&=\partial_\mu\phi+ig(A_\mu * \phi -\phi * A_\mu)\cr
 &= \partial_\mu \phi+ig\left[A_{\mu}(x-{L\over 2}) \phi(x) -
\phi(x) A_{\mu}(x+{L\over 2})
\right]\ ,}}
and as usual the $\phi$ field under gauge transformations
should transform as
\eqn\adjGT{
\phi \to U*\phi* U^{-1}\ .}
As it is seen from \adjoint if we start with $SU(N)$ gauge group, i.e.
$\phi, A_\mu \in su(N)$, in the interaction vertices we will have both
the completely symmetric and anti-symmetric $su(N)$ tensors (usually
denoted by $d^{abc}$ and $f^{abc}$ respectively). The appearance of
$d^{abc}$ factors, to make the theory ``renormalizable'',  eventually will
force us to include the central $U(1)$ factor. In other words, in the
noncommutative dipole theory with the adjoint matter, it is not possible
to have a $SU(N)$ theory while $U(N)$ is possible. We would like to note
that this case is actually what we find from string theory.

Hereafter we restrict ourselves to the $U(1)$ case.
The dipole moment of the adjoint matter with dipole length $L$, as we
expect, is twice bigger than the fundamental matter. It is worth
noting
that in this case, since the dipole is lowest pole present (there is no pure
charge), the field redefinition through simple translation does not work.
One can check that re-defining $A_{\mu}(x-{L\over 2})-A_{\mu}(x+{L\over
2})$ as the new gauge field,  we will lose the simple gauge
transformations rule for the re-defined gauge field. (Or equivalently
the gauge theory action will not look as simple as
$F_{\mu\nu}F^{\mu\nu}$.)

The fermion-photon vertex in this case is obtained by replacing the
exponential noncommutative dipole factor of the fundamental matter, by
$2i\sin {1\over 2}p\cdot L$. Here we present the results of  loop
calculations and the  renormalizability for the $U(1)$ case,
and the full
and detailed study of the noncommutative dipole theories with adjoint matter
is postponed to future works \refs{\future}. One should note that for the
noncommutative
dipole QED (with adjoint matter), the corresponding commutative theory
is trivial, it is just an uncharged particle plus a  pure $U(1)$ theory.
\vskip .3cm
\noindent
{\it One loop photon propagator}
\vskip .3cm
\noindent
Upon insertion of the dipole sine factor, the photon self-energy diagram
is multiplied by the factor of $-4\sin^2({1\over 2}p\cdot L)$ (times the
usual QED result coming from the fermion running in the loop). In other
words the noncommutative dipole factor will not enter into the loop
integral. So, the divergent part of this diagram will show the same tensorial
structure as the usual QED and hence the gauge invariance is guaranteed.
However, because of this sine factor now the renormalization factor for
the photon field, $Z_A$, is  multiplied by a factor of $-4\sin^2{{1\over
2}p\cdot L}$. In other words, to absorb the divergences, besides the
usual $F_{\mu\nu}F^{\mu\nu}$ type term we should add
$F_{\mu\nu}(x-{L\over 2}) F^{\mu\nu}(x+{L\over 2})$ type counter-terms.
At higher loop level still the counter-terms needed, can be expressed in
terms of $F_{\mu\nu}$, but in a more non-local way (similar to the
\counterterm).

\vskip .3cm
\noindent
{\it One loop fermion propagator}
\vskip .3cm
\noindent
In the noncommutative dipole QED (with adjoint matter),
there is only one diagram which contributes
to the one loop fermion two point function. Here the noncommutative dipole
factor will enter into the loop integrals, i.e. we have planar and non-planar
diagrams. The planar part is essentially the same as usual QED, while the
non-planar part is finite and there is no IR/UV mixing.
\vskip .3cm
\noindent
{\it One loop fermion-photon vertex}
\vskip .3cm
\noindent
The only diagram that contributes to the fermion-photon vertex contains both
planar and non-planar diagrams. Again (up to some numeric factors) the
divergent part (coming from planar diagram)  is the same as usual QED, and
the non-planar part is finite.

All together, the above mentioned theory is a sensible theory (in the
same spirit as discussed in sub-section ${\em 3.1}$), however the
$\beta$-function is different from the usual QED. The theory is in fact
asymptotically free (note that there is an extra factor of $i$ in the
expression for the vertex). Moreover,  unlike the Moyal  noncommutative gauge
theory where the noncommutativity parameter $\theta$ do not appear in the
$\beta$-function explicitly, in the noncommutative dipole case, dipole length
$L$ will enter into the expression for the $\beta$-function.  However, the
dipole length itself is not receiving any quantum corrections.

\section{Supersymmetric Noncommutative Dipole Gauge Theory}

Having defined noncommutative dipole gauge theories, one can check that
it admits a supersymmetric extension, though the maximal supersymmetric
case  is now  ${\cal N}=2$ in four dimensions (8 supercharges).
To see this we recall that, as discussed earlier,
the gauge fields should appear with zero dipole lengths and supersymmetry
requires that all the fields in the vector multiplet to have the same
dipole length. Therefore the $D=4$, $\cN=4$ case which only contains
the vector multiplet does not admit a non-trivial dipole extension.  \foot{
We would like to comment that this is not the case with the Moyal
noncommutative gauge theory, which admits 16 SUSY extension. In that case
using the Seiberg-Witten map, the noncommutative theory, can be understood
as the commutative theory perturbed by a  dimension six (and higher
dimensional) operators which all preserve $SU(4)$ R-symmetry \refs{\Ken}.}
The $\cN=4$ vector multiplet, in the $\cN=2$ language can be decomposed
as a vector multiplet + a hyper multiplet. So, the first possibility
arises when we allow the $\cN=2$ hyper multiplet,
which is of course in the adjoint representation of the gauge group,
to have a non-zero dipole length (while the vector multiplets still have
zero dipole lengths).  {}From the $\cN=2$ point of view, the dipole length
of the hypermultiplets can be understood as a new (dimensionful) moduli of
the theory. In other words, the dipole version of $\cN=4$ can be
understood
as a perturbation of the commutative $\cN=4$ case by dimension five (and
higher dimensional) operators.  Such a perturbation will break the
$SU(4)$ R-symmetry. For the case that the two chiral multiplets in the
$\cN=2$ hyper multiplet
have the same dipole length, the R-symmetry is
broken to $SU(2)_R\times SU(2)_L \times U(1)$, i.e. an $\cN=2$ theory.
We note that the specific $\cN=2$ theory described above, in the zero
dipole length limit goes back to a $\cN=4$ theory and therefore, this theory
may be understood as a noncommutative dipole extension of $\cN=4$ gauge
theory.
In fact, in the next section we will present the brane configuration which
exactly leads to this noncommutative dipole SYM theories.

Of course one can consider the theories with lower SUSY, by
further breaking of the R-symmetry. This is possible by assigning
different dipole  lengths to the chiral matter fields.

\section{String Theory Realization of Noncommutative Dipole Theory}

In this section we will study how to embed the noncommutative dipole
theories in string theory. This issue has been discussed earlier
in \refs{\BDGKR,
\DGR}. When we orient the background $B_{NS}$ field so that it has
only one leg parallel to
the brane then the low energy theory on the world volume is a
noncommutative dipole theory.

First we show how the dipole star product \prodrule arises in the twisted
compactified string theory. Then we present the supergravity solutions
corresponding to T-dual of D2-brane with a twisted compactification
\refs{\BDGKR}. Here we will concentrate on a special twist which preserves
eight supercharges.  Taking the near horizon limit allows us to study the
UV and IR behaviour of the large $N$ limit of  the supersymmetric
noncommutative dipole gauge theories.
As an alternative  way, we study  a D3-brane probing a Taub-NUT
with a particular $B_{NS}$ background  \refs{\DGR} and discuss the
similarities and differences between the two approaches.

\subsection{Dipole star product from twist}

Consider a D2-brane along $012$ directions with the following ``twisted''
compactification
\eqn\dthreeprobe{
(x_{012}, x_3, x_{3+a}) \to (x_{012}, x_3+2\pi R,
\sum_{b =1}^6 O_{ba}x_{3+a}), }
where $a=1,...,6$. The explicit form of $O$ is given by
$O = e^{2\pi i R M \over \alpha'}$ where $M$ is a finite matrix of the Lie
algebra $so(6) \equiv su(4)$ with dimensions of length. Under the action of
$O$
\eqn\xa{ \delta x_a = \Omega_{ab} x_b dx_3 ,}
where $\Omega = 2\pi i M/\alpha'$.
The rank of $\Omega$ determines the number of supersymmetry preserved. In
\refs{\BDGKR} $\Omega$ was of rank 4 and therefore the model had no
supersymmetry. \foot{ The Melvin twisted compactification of Matrix Models
have also been discussed in \refs\Motl .} We shall take $\Omega$ with rank 2
preserving ${\cal N}=2$ SUSY.
The low energy effective field theory of the above brane configuration,
after a T-duality in third direction, is a noncommutative dipole theory.
To see this, we review and expand the arguments of \refs{\BDGKR}  showing
how the dipole product can be derived from the above twist analysis.
First we note that  the dipole product of \prodrule can be
equivalently written as
\eqn\dipprod{
(\phi_1 * \phi_2)(x) \equiv exp {\Big (}{1\over 2} (L_1^\mu{\del
\over \del x_2^\mu} - L_2^\nu {\del \over \del x_1^\nu}) {\Big )}
\phi_1(x_1)\phi_2(x_2){\Big \vert}_{x_1=x_2=x}}
which tells us that inserting a phase
\eqn\phase{
e^{i \sum_{1\leq i <j \leq n} p_i\cdot L_j}}
with the requirement $\sum_{i=1}^n L_i = 0 = \sum_{j=1}^n p_j$, \foot{
The necessity of these requirements from  the
noncommutative dipole field theories side have been discussed earlier.}
in front of terms like
\eqn\termslike{
tr[\phi_1(p_1)\phi_2(p_2)....\phi_n(p_n)]\ , }
will generate the noncommutative  dipole theory from a given commutative
theory.

To show how the phase factor \phase arise from the twist analysis, we
consider D2-branes probing the
twisted background \dthreeprobe . For simplicity
we can assume the twist acting on only two
coordinates $Z=x_6+ix_7$ as $Z \to e^{i\alpha} Z$, in terms of our previous
notation, $\Omega$, $\alpha = \Omega R$.
In the presence of the twist,
going round the compactified circle, open strings can gain a non-trivial
``winding'' number along the $Z$ directions. This winding is then an
integer multiple of $\alpha$, the twist angle. Following \refs{\BDGKR}, we
call this winding as $Z$-charge. Now, let us consider
$n$ interacting open strings with winding numbers $w_1, ..., w_n$ (around
the $x_3$ direction) and with $Z$-charges $q_1, ..., q_n$. The scattering
amplitude for these $n$ open strings is equal to the zero twist case,
but, now we should also insert the total
phase of
\eqn\totphase{
e^{i\sum_{1\leq i <j \leq n} w_i q_j}\ .}

We would like to stress that to have a meaningful twisted
compactification, we should confine the scattering amplitudes to
those which  respect the symmetries need for twisted compactification of
\dthreeprobe.
This means that in our twisted string theory only the scattering
processes with the vanishing total windings are allowed. In other words,
besides
inserting \totphase phase factor we should also restrict them to
\eqn\condition{
\sum_{i=1}^n w_i=0\ ,\ \ \ \ \ \sum_{i=1}^n q_i=0\ .}
Now, we make  a T-duality in $x_3$ direction according which, we should
replace $w_i$ with the momenta in third direction and the $Z$-charges
with the dipole length $L_i$ (more precisely,
$L_i=\tilde R q_i$, where $\tilde R={\alpha'\over R}$ is the
compactification radius after T-duality). So upon the T-duality, \totphase
will produce the looked-for phase factor of \phase . It is worth noting that,
all the dipole vectors obtained in this way are along the T-duality
direction, $x^3$; moreover the ratio of the dipole length for different
fields can only be a rational number. We would like to point out that
in order to find finite dipole lengths in the $\tilde R\to \infty$
(decompactification) limit  one should send the twist angle, $\alpha$ to
zero. However, this is compatible with the gravity
decoupling limit, which we will discuss in the next part.

However, one should note that in the low energy effective field theory of
the above twisted open strings, all the possible $q_i$ should be
considered, i.e. if the lowest dipole lengths for a twisted string is
denoted by $L$, a field can have all the integer multiples of this dipole
length. More precisely, this low energy theory is a special case of
noncommutative dipole theory we introduced and studied in previous
sections. For example, in the $\phi^4$ case, in fact the potential that we
find from the above analysis is
\eqn\stringdipole{\lambda\sum_{Q=0}^{\infty}(\phi^{\dagger}\phi)(x-{QL\over
2})
(\phi^{\dagger}\phi)(x+{QL\over 2}).}
If we just ignore the higher $Z$-charges, and start with the lowest
dipole lengths, these terms will be generated
through loop effects. {}From
the field theory point of view this means that in order to regularize the
theory, although we do not require new degrees of freedom,  we need to
add some non-local counter-terms; i.e. if we start with a dipole length
corresponding to $q_i=\alpha$ as the lowest dipole length in the above
notation, then the non-local counter-terms would involve  all the integer
multiples of $\alpha$. Note that a twisted string with multiple windings
should not be treated as a new degree of freedom (a new field).


\subsection{Supergravity background}

The supergravity solution which shows a D3-brane probing the above twisted
geometry can be obtained by starting with a D2-brane and making a
T-duality in the third direction \refs\BDGKR. Noting the definition of the
twist, the ``genuine'' distance scale
is $dx_a - \delta x_a$ which gives rise to the following background
metric
\eqn\metricnow{
ds^2 = f^{-1/2}{\Big (}dx^2_{012}- {dx_3^2\over  1+(\Omega x)^2}{\Big )}
- f^{1/2} {\Big (} dx_adx_a - {(dx. \Omega x)^2\over 1+(\Omega x)^2}{\Big
)}\ ,}
In the above we have made $x_3$ dimensionless by putting $R = l_s = 1$
where $R$ is the radius of the compact $x_3$ direction.  The twist $\Omega$
also generates a $B_{NS}$ field given by
\eqn\bfieldfrtwist{
B_{3a}dx_a = -{\Omega x . dx \over 1+(\Omega x)^2}\ .}
It is worth noting that under the four dimensional parity transformation
the above B-field will change sign. (This is not the case with the Moyal
case where $B$ has two legs along the brane and hence does not change sign
under parity.) So, the noncommutative dipole theory that we obtain in
presence of the above B-field is parity violating.

Since the noncommutative dipole theory is a decoupled theory the near
horizon geometry
determines the gravity dual of the theory. In fact the decoupling limit is
obtained by $\alpha'\to 0$ while keeping $\Omega$ fixed. In terms of $\Omega$
the dipole lengths $L\sim \alpha' \Omega Q$, where $Q$ is an integer
determining the winding along the twisted direction. Therefore, the
non-trivial dipole effects only appear from the large  $Z$-charges, or
in the \stringdipole through the terms with large $Q$ ($\alpha'
Q=fixed$).  Writing $x_a= rn_a$ with
$\|n_a \|^2=1$,
the near horizon metric is
\eqn\nearhori{
ds^2 = {1\over u^2}{\Big (}dx^2_{012}-du^2- {u^2\over u^2+(\Omega n)^2}
dx_3^2 {\Big )} - {\Big (} dn^2 - {(\Omega n.dn)^2\over u^2+(\Omega n)^2}
{\Big )}\ ,}
where $u = 1/r$.
Observe that the metric is a deformed AdS and a deformed $S^5$. Deformed
AdS shows that the theory is no longer conformal which is expected because
there is an inherent scale $-$ the dipole length $-$ in the theory. Deformed
sphere tells us that the R symmetry is no longer $SO(6)$. In fact from the
metric its clear that, taking a rank 2 twist, the R-symmetry is $SU(2)
\times SU(2) \times U(1)$.
The behaviour of the
NSNS two form is given by
\eqn\twoform{
\sum_a B_{3a}dn_a = -{\Omega n.dn \over u^2 + (\Omega n)^2}\ .}
In the near horizon this field breaks Lorentz invariance explicitly on the
world volume theory. The dilaton behaves as
\eqn\dilatonbeh{
e^{2(\phi-\phi_0)}= {u^2\over u^2 + (\Omega n)^2}\ .}

\noindent (a) {\it IR physics}

The parameter $u$ determines the RG scale of the theory. Therefore the IR
corresponds to large $u$. For large $u$ the background is $AdS_5 \times S^5$
and hence the noncommutative dipole theories are determined from SYM by
perturbing it by a
dimension 5 operator \refs{\BerGan, \DGR}
\eqn\dimfive{
{i\over g^2_{YM}}{\Big (} tr[F_{\mu\nu} \phi^{[I}D_{\nu}\phi^{J]} +
\sum_K (D_{\mu})\phi^{[K}\phi^I \phi^{J]}] + fermions{\Big )}\ ,}
where $I,J=1,..,6$ are R-symmetry indices, $\phi^I$ are the scalars, $D_{\mu}$
is the covariant derivative with respect to gauge fields $A_{\mu}$ of
field strength
$F_{\mu\nu}$ and $[....]$ is complete antisymmetrization. The operator \dimfive
transforms in the ${\bf 15}$ of the R-symmetry group $SU(4)$. As expected, the
dual of this operator in supergravity is our rank two field $B_{NS}$
transforming as ${\bf 15}$. The other correspondences from supergravity
for the field theory at far IR have been worked out in \refs{\BDGKR}.

\noindent (b) {\it UV physics}

The UV physics appears on the gravity side when we go near the boundary or
equivalently
when $u\to 0$. Something interesting happens now. {}From the form of the
metric we see that some of the components vanish. In fact the fibered
circle of the
deformed $S^5$ $-$ which is a $S^1$ fibered over a base $CP^2$ $-$ shrinks
to zero size. T-dualising along that direction we obtain the metric
\eqn\metiniia{
ds^2= {1\over u^2}[ds^2_{012u}-(dx_3+ Ld \gamma)^2] - d\gamma^2 -
ds^2_{CP^2}\ ,}
where $\gamma$ is the circle coordinate. The dilaton is a constant and
there are non-trivial $H_{NS}$ and four-form backgrounds.

The above form of metric makes explicit the non-local nature of the
underlying field theory:

\noindent (i) The proper distance between two
points $(x_3,\gamma)$ and $(x_3+2\pi L,
\gamma)$ is $2\pi$ which is of stringy scale (we have taken $R = l_s =1$).

\noindent (ii) The proper distance between two points $(x_3,\gamma)$ and
$(x_3+ \Delta,\gamma)$ when $\Delta$ is not an integer multiple of $L$ is
${1\over u} \to \infty$.

This should be compared with the similar case that we observe for the
noncommutative Moyal theory. The metric along the noncommutativity directions
$x_2, x_3$ goes as \refs{\aki,\juan}
\eqn\metfornoncom{
ds^2 = {u^2\over u^4 + \theta^4}(dx_2^2+dx_3^2)\ ,}
where $\theta$ is the noncommutativity parameter. At the UV, i.e when
$u \to 0$, the metric shrinks to zero. Assuming $x_{2,3}$ forming a torus,
\foot{This will avoid the IR problems also.}
we T-dualise the above metric to get
\eqn\ncgiia{
ds^2 = {1\over u^2} (\theta dx_2 + dx_3)^2\ ,}
which is of the same form as \metiniia. Also observe that
from
\metiniia we actually {\it restore} the 4 dimensional superconformal
invariance.
This was also seen from some related field theory loop calculations in the
earlier sections. Before we end we note that the problem regarding T-duality
and fermions which appeared in \refs{\BDGKR} because of the non-SUSY
background, will not appear here.

\subsection{Noncommutative dipole theories from Taub-NUT background}

In \refs{\DGR} another way of studying noncommutative dipole theory was
developed. It was shown that when we place a D3-brane near the origin of a
Taub-NUT space and switch
on a $B_{NS}$ background which has one component along the D3-brane and
another component along the shrinking cycle of the Taub-NUT the low
energy theory on the world volume of D3-brane  is a noncommutative dipole
theory. Below we mention the steps.

Consider
a D3-brane oriented along $x_0, x_1,x_2,x_3$ and orthogonal to a
Taub-NUT space along $x_6,..., x_9$.
In the absence of a $B$
field the metric of a Taub-NUT space probed by a D3-brane
is non-singular in a good coordinate system $u'$, and is given by
\eqn\dprobe{
ds^2=f_2^{-1/2} ds^2_{0123}+ f_2^{1/2}[ds^2_{45}+ du'^2+ u'^2 d\Omega^2
+ u'^2(dx_6 + B_{6i}dx_i)^2]}
with $f_2^{-1/2} = u'^2$ in the near horizon limit.

Let us now switch on a  $B_{NS}$ field with one leg oriented along the
brane. If the asymptotic value of the $B_{NS}$ field is $b$ and
\eqn\hdef{
h^{-1} = sin^2\theta + f_1 cos^2\theta}
then the classical SUGRA background for the system is given by
\eqn\classback{
ds^2= f_2^{-1/2}[ds^2_{012}+hf_1dx_3^2]+f_2^{1/2}[ds^2_{45}+du'^2+  u'^2
d\Omega^2 + h(dx_6+B_{6i} dx_i)^2]}
where $f_1 = u'^{-2}$ in the near horizon limit.
The above metric goes to a flat one asymptotically. Using the good
coordinate system $u'$, the D3 probes a smooth Taub-NUT space with a
$B_{NS}$ field. Therefore from the above
analysis
we expect that the metric component $g_{33}$ in \classback is given, for small
values of $b$,
by
\eqn\gthree{
g_{33} = hf_1 f_2^{-1/2} = {u'^2\over 1+ u'^2 b^2}}
To compare this to \nearhori we have to identify $u= 1/u'$. Under this
identification the $g_{33}$ components look similar {\it if} we replace
$\mid \Omega n \mid $ with $b$. To see whether
this is a generic phenomena with our
background we have to identify the other components of the metric and the
NSNS field. The background $B'_{NS}$ field is given by
\eqn\bfieldback{
B' = h~tan\theta~ dx_3 \wedge (dx_6 + B_{6i}dx_i) =
{b\over b^2+u'^{-2}}~dx_3 \wedge (dx_6 + B_{6i}dx_i)}
This is the same as \twoform if we replace $\mid \Omega n \mid $
with $b$. Now let us
check for the metric along the sphere direction. The components $g_{66}$ and
$g_{77}$ (and also $g_{67}$) have coefficients
\eqn\coef{
h~f_2^{1/2} = {1\over 1+u'^2 b^2} = {u^2\over u^2 + b^2}}
which is again same under the above replacement. Finally its easy to check
that the dilaton also behaves in the expected way.

One last thing to identify is the nature of the dipoles in the
noncommutative dipole theory. This can be argued in two ways. First is directly
from
D3 probing the Taub-NUT with $B_{NS}$.
In \refs{\DGR} the dipoles in these theories were identified with rotating
arched strings. These strings have angular momentum along (say) $x_6, x_7$
directions and they rotate in the background three form field $H_{367}$. The
D3-brane, as usual, is stretched along $x_{0123}$. The system is stable
because the tension of the string is balanced by the outward pull due to the
rotating string in $H$ background. The identification is now clear from the
fact that R-symmetry corresponds to the simultaneous rotations of the
$6-7$ and $8-9$ planes by the same angle. The dipole length in this case will
be determined by the angular momentum. The fact that interactions of these
strings should preserve the {\it internal} angular momentum implies that
$\sum_i L_i = 0$. However this simple classical picture is valid in the
limit of large $b$. Otherwise we have to take into account the radiative
corrections for the metric and the $B$-fields radiations.

The second is from the T-dual version of the above model.
As we know, the T-dual picture is a NS5-brane oriented along $x^{0,1,...,5}$
and a D4-brane along $x^{0,1,2,3,6}$ where $x^6$ is the compact direction. The
directions $x^3, x^4$ are on a slanted torus and therefore the D4 comes back to
itself with a shift along $x^3$. The strings connecting the D4 across the
NS5-brane are expected to form dipoles and
are charged under $(N, {\bar N})$ of the gauge group $U(N)$. Therefore they
transform as an adjoint. Its also clear from the brane construction that
the vectors do not pick up a dipole length.
 Recall that this is what we need from
our earlier discussions on adjoint matters. However if the matter is
in the fundamental then we expect product gauge groups to have a non-trivial
noncommutative dipole theory. This also
can be easily
seen from multiple parallel NS5-branes on the $x^6$ circle with D4-branes
along that circle.
In the original
model this corresponds to a multi Taub-NUT probed by D3-branes.
The non-conformal extension of this is to
probe the background with integer and fractional D3-branes. Existence of
fractional D3 branes gives rise to a scale in the theory. And therefore there
would be logarithmic RG flow.
Switching on further
$B_{NS}$ fields will generate dipole lengths of the fundamental matters on
the D3-brane with product gauge groups. This would then be a realization of
the noncommutative dipole theory with a RG flow.

Therefore from the above detailed analysis we can conclude that the
noncommutative dipole
theory generated using a Taub-NUT background is an approximation of the
one generated using a twist (when we take the twist metric to have a
rank two). This is encouraging because the Taub-NUT background is useful to
do BPS analysis and now under this identification it can therefore be
extended to the twisted case too. However this simple identification
doesn't extend (as far as it seems) to higher rank twists.

\section {M-Theory Realization}

As discussed in the introduction, the oriented $B_{NS}$ fields give
various new theories. These theories can also be realized from M-theory
by keeping a M5 brane near a Taub-NUT singularity and switching on an
appropriate $C_{\mu\nu\rho}$ field. The M5-brane is oriented along
$x^{0,1,2,3,4,5}$ and is orthogonal to a Taub-NUT space along $x^{7,8,9,10}$
where $x^7$ is the Taub-NUT circle.
The noncommutative dipole theory can be generated from
a M5 brane with a $C$ field having two components $\mu,\nu= 4,5$
  along the M5-brane and
the other component $\rho$ along the Taub-NUT circle $x^7$.

When the external parameters are carefully chosen this leads to a decoupled
noncommutative dipole theory in six dimensions. The limits of the external
parameters are:
\eqn\decoup{
C \to \epsilon,~~~
R_7 \to \epsilon,~~~
M_p \to \epsilon^{-\beta},~~~ \beta >1}
In this limit the energy scale of the excitations of the M5-brane is
kept finite whereas the other scales in the problem are set to infinity.
This decoupling is kinematical. For a different scaling of external
parameters
\eqn\limits{
C \to finite,~~~
R_7 \to finite,~~~
M_p \to \infty,~~~R_{10} \to 0}
we get a dynamical decoupling. This decoupling is in the same spirit as
the little string theory.

However in both the cases above we have kept the value of $C$ very low.
An interesting case is when $C\to \infty$ and we remove the M5-brane
from the picture.
It turns out that if we
consider the following limits:
\eqn\limittwo{
C \to \infty,~~~M_p \to \infty,~~~M_p^3 C^{-1} \to fixed}
with the identification that the M-theory circle is now $x^7$ we get a
$6+1$ dimensional noncommutative YM theory whose coupling
constant \eqn\cop{g^2_{YM} = M_p^{-3}C = fixed}

This limit is consistent with (and in fact it's the same as) the limit
studied by Seiberg-Witten.
The exact limits which give us a noncommutative $6+1$ dimensional theory
is:
\eqn\anot{
C \to \epsilon^{-1/2},~~~
M_p \to \epsilon^{-1/6},~~~
R_7 \to constant, ~~~
g^M_{\mu\nu} \to \epsilon^{2/3}}
where $g^M_{\mu\nu}$ is the dimensionless M-theory metric.
However, as discussed in \refs{\AOI}, unlike what is naively expected,
the theory is not decoupled from gravity.

\section{Other Kinds of *-Products}

In the above sections we discussed about the new kind of *-product coming
from noncommutative dipole theory. This *-product, as opposed to
noncommutative theory,
doesn't affect the underlying space-time. But as seen from the supergravity
point of view, these theories share the non-local
behaviour at the UV region.

An interesting extension is the idea of {\it nonabelian geometry}
\refs{\KesYin}.
 The
derivation of the nonabelian *-product takes into account both the
noncommutativity of space-time and the dipole nature of the open strings.
The two ends of these dipoles lie on noncommutative spaces with different
fluxes on them. Interaction of these dipoles can be thought of as interactions
between the center-of-mass of these dipoles.

The end points of the dipoles are labelled by $x_1$ and $x_2$ respectively
and each of the end points see different noncommutativity parameter
$\Omega_i$ and $\Omega_j$. These $\Omega$'s can be transformed as
canonical forms
\eqn\canform{
\Omega_i = T_i J T^{\top}_i}
where $J$ is a canonical matrix \refs{\KesYin}. {}From the above
decomposition
the center-of-mass of our dipole turns out to be:
\eqn\com{
x_c \equiv (T_i^{-1}+ T_j^{-1})^{-1} (T_i^{-1}x_1 + T_j^{-1} x_2)}
This choice of center-of-mass changes the interaction matrices and therefore
affects the *-product for this theory as:
\eqn\newstar{
(\Psi^i_j *_{ijk} \Phi^j_k)(x) =  \exp {\paren {\frac \imath 2
		\frac \pa {\pa{x'^{\mu}}}\Omega_{ij;jk}
		 \frac \pa {\pa{x''^{\nu}} }}}
          \Psi^i_j(x')\Phi^j_k(x'')
          \Big\vert_{x'=S^{ik}_{ij}x,x''=S^{ik}_{jk}
          x}}
where $S$ is defined in \refs{\KesYin} and $\Omega_{ij;jk}$
can be given in terms of $T_i$ as:
\eqn\defofomega{
\Omega_{ij;jk}= \left( {T^{-1}_i+T^{-1}_j \over 2} \right) ^{-1}J~
\left( {T^{-1}_j+
      T^{-1}_k \over 2}\right) ^{{\top}.{(-1)}}}
The above *-product \newstar
can be interpreted as the star product for multiple
branes having different noncommutativity parameter on them.
The quantity $J$
 is the usual noncommutativity
\foot{Multiple noncommutativity on the branes
were first discussed in \refs\RT, in connection with conifold geometry, and
in \refs\DolNap, for parallel branes in a slowly varying background field.}
(due to $B$ field) and $T_i$'s
are due to the non abelian nature. \defofomega\ therefore encodes this
intertwining clearly and there is no way to separate them \refs{\KesYin}.

\subsection{Embedding in String Theory}

From the above constructions of the nonabelian theory it would seem difficult
to have supersymmetric brane configurations with different fluxes on their
world volumes.
However there are two interesting configurations which {\it require}
different fluxes on their world-volumes to be stable:

\noindent (i)A pair of $D5 -{\bar D5}$ wrapped on
two cycle of a conifold. The absence
of tachyons in this system require different fluxes $b_i$ on
their world-volumes. The zero point energy of a string connecting them is
\eqn\zeropoint{
E = -{1\over 2} {\Big (} \vert \nu - {1\over 2} \vert + {1\over 2} {\Big )}}
with $\nu$ being the shift in the mode number given by
\eqn\shift{
e^{2\pi i\nu} = {(1-ib_1)(1+ib_2)\over (1+ib_1)(1-ib_2)}}
Using the right GSO projection one can show that the tachyon in this system
becomes massless. For this system one can show that the *-product
simplifies as
\eqn\simpleprod{
\psi_{ij} *_{ijk} \phi_{jk} \approx \psi_{ij} *_{j} \phi_{jk}}
where $*_j$ is defined in terms of
$\Omega_j \equiv \Omega_{jj;jj}$.

\noindent (ii)A pair of $Dp - {\bar Dp}$ in flat
space with world-volume magnetic and
 electric fields turned on. This system is ${1\over 4}$ BPS because in the
presence of these fields the SUSY preserved by both branes and anti-branes
are same as long as the magnetic fields come with opposite sign. This
construction has been discussed in details in \refs{\bkar}.

\section{Discussions and Open Problems}

Motivated by string theory brane configuration, where a D3-brane is probing
the ``twisted'' geometry of \dthreeprobe or a D3-brane probing a Taub-NUT space
with a B-field turned on, we studied the noncommutative dipole field theories
(NCDFT's) in more detail. Besides the string theory, there are
some strong  motivations from the usual particle physics: charged leptons,
neutrinos and neutron and proton have non-zero electric dipole moments
\refs{\ParticleData}.  Furthermore the existence of an inherent electric
dipole moment is a sign of CP violation. In fact, as we discussed,  the
noncommutative dipole gauge theories we have studied here in some particular
cases show signs of CP and even CPT violation.

With the above motivations we introduced the $C^*$-algebra of functions
with the appropriate dipole star product \prodrule. Then,  studying
the renormalizability of noncommutative dipole theories, we showed that there
is no IR/UV mixing in this case.

As the first open question to mention here, it would be very interesting to
build up a full dipole version of the Standard Model. In this case, unlike
the Moyal case \refs{\NCSM}, it is possible to have $SU(N)$ noncommutative
dipole theories with the fundamental matter fields,
and building such a model is more straightforward. Furthermore, we do not
face charge quantization problem of the Moyal noncommutative QED
\refs{{\Haya},{\NCSM}}. In the dipole version,
although the Lorentz symmetry is broken, we have the advantage that
the theory is still under control and we have a sensible field theory.
In particular, we would like to mention the possibility to  consider
the neutrinos as an adjoint matter field under the dipole version of QED.

The other interesting problem we would like to address here is solitonic and
topological solutions of the NCDFT's. As we discussed , the dipole star
product will not appear in the dipole- scalar and pure gauge theories and
hence here we do not find the solutions of Moyal noncommutative
theories discussed in Ref.\GMS . However, the noncommutative dipole
version of the 't Hooft-Polyakov monopole solutions will keep traces of
the dipole deformation. It is straightforward to show that the BPS
monopole equations in the dipole case reads as
\eqn\BPS{
\vec{B}^a(x-{L\over 2})=\pm(\vec{D}\Phi^a) (x)\ ,}
where $B_i^a=\epsilon_{ijk} F^{ij a}$,  $a$  stands for the
SU(2) indices and $\pm$ corresponds to monopole (or anti-monopole) solutions.
Then one can check that the solution to the above equation is obtained
by making the shift, $x\to x+{L\over 2}$ in the argument of the gauge field.
It is readily seen that  the monopole charge for this solution
($\int_{S^2_{\infty}} \Phi^aB^a\cdot dS$) would give the same value as
the commutative case. However, this monopole solution
besides the monopole charge, also carries (magnetic) dipole moment.

{}From the string theory point of view the noncommutative dipole theories
fall under the general framework of studying various theories on the
world-volume of
D3-branes with $B_{NS}$ fields oriented in different ways on the D3-branes.
As it is studied in detail here and also in \refs{\DGR,\BDGKR} when we
have
a $B_{NS}$ field with one leg along the D3 brane and other leg orthogonal to
it, the low energy theory on it is a NCDFT.
However when we orient the $B_{NS}$ field completely orthogonal to the
D3-brane the world volume theory is the {\it pinned brane}
theory \refs{\CDGR}.
Here the D3 has minimum tension at some point in space (origin of a
Taub-NUT singularity). The hypermultiplet in these theories are massive.
Finally, as is well known, when we have both the legs of the $B_{NS}$ field
along the brane the world volume theory is noncommutative geometry.

\noindent {\bf Acknowledgements}

We would like to thank Savas Dimopoulos,
Ori Ganor, David Kaplan, Mark Van Raamsdonk and Lenny
Susskind for many helpful discussions, M. Alishahiha for
comments on  the manuscript and the Stanford theory group for critical
comments. The research of K.D. is supported
in part by David and Lucile Packard Foundation Fellowship 2000-13856. The
research of M. M. Sh-J. is supported in part by NSF grant PHY- 9870115 and
in part by funds from the Stanford Institute for Theoretical Physics.

\listrefs
\end